\documentclass{jdsart}
\usepackage{bm}
\usepackage{mathtools}

\articletype{research-article}
\firstpage{1}
\lastpage{11}

\theoremstyle{plain}

\theoremstyle{remark}

\theoremstyle{definition}

\begin{document}
\begin{frontmatter}

\title{Stellar Blend Image Classification Using Computationally Efficient Gaussian Processes}
\runtitle{Efficient Gaussian Processes}

\author[1]{\inits{C.}\fnms{Chinedu} \snm{Eleh}\thanksref{f1}
\ead{cae0027@auburn.edu} \thanksref{c1}
}
\author[2]{\inits{Z.}\fnms{Yunli} \snm{Zhang}\thanksref{f1}}
\author[3]{\inits{R.}\fnms{Rafael} \snm{Bidese}\thanksref{f1}}
\author[4]{\inits{P.}\fnms{Benjamin W.} \snm{Priest}}
\author[5]{\inits{M.}\fnms{Amanda L.} \snm{Muyskens}}
\author[1]{\inits{M.}\fnms{Roberto} \snm{Molinari}}
\author[1]{\inits{B.}\fnms{Nedret} \snm{Billor}}

\thankstext[id=f1]{Authors contributed equally.}
\thankstext[type=corresp,id=c1]{Corresponding author.}
\address[1]{Department of Mathematics and Statistics, \institution{Auburn University}, \cny{Auburn, AL, USA}}
\address[2]{Department of Mechanical Engineering, \institution{Auburn University}, \cny{Auburn, AL, USA}}
\address[3]{Department of Biosystems Engineering, \institution{Auburn University}, \cny{Auburn, AL, USA}}
\address[4]{Center for Applied Scientific Computing, \institution{Lawrence Livermore National Laboratory}, \cny{Livermore, CA, USA}}
\address[5]{Computational Engineering Division, \institution{Lawrence Livermore National Laboratory}, \cny{Livermore, CA, USA}}

\begin{abstract}
Stellar blends, where two or more stars appear \textit{blended} in an image, pose a significant visualization challenge in astronomy. Traditionally, distinguishing these blends from single stars has been costly and resource-intensive, involving sophisticated equipment and extensive expert analysis. This is especially problematic for analyzing the vast data volumes from surveys, such as Legacy Survey of Space and Time (LSST), Sloan Digital Sky Survey (SDSS), Dark Energy Spectroscopic Instrument (DESI), Legacy Imaging Survey and the Zwicky Transient Facility (ZTF). 
To address these challenges, we apply different normalizations and data embeddings on low resolution images of single stars and stellar blends, which are passed as inputs into machine learning methods and to a computationally efficient Gaussian process model (MuyGPs). MuyGPs consistently outperforms the benchmarked models, particularly on limited training data. Moreover, MuyGPs with $r^\text{th}$ root local min-max normalization achieves 83.8\% accuracy.  Furthermore,    MuyGPs' ability to produce confidence bands ensures that predictions with low confidence can be redirected to a specialist for efficient human-assisted labeling.

\end{abstract}


\end{frontmatter}

\section{Introduction}
The Vera C. Rubin Observatory is set to revolutionize our understanding of the cosmos with its upcoming Legacy Survey of Space and Time (LSST)  with first light expected in 2025 \cite{ivezic2019lsst, rubin_2024}. The LSST design is a large, wide-field ground-based system, driven by four scientific purposes: probing dark energy and dark matter, taking an inventory of the solar system, exploring the transient optical sky, and mapping the Milky Way. It aims to comprehensively investigate various scientific goals, such as weak lensing shear. This phenomenon involves the correlated yet subtle distortion of multiple galaxy images, a result of gravitational lensing caused by a shared concentration of intervening matter \cite{buchanan2022gaussian}. 
Gravitational lensing plays a vital role in demystifying dark matter and dark energy by analyzing the distortion of light from celestial objects as it passes near massive objects, providing insights into the universe's unseen matter. However, this method faces challenges from stellar blending, where stars within the same line-of-sight (LOS) appear as a single entity in images, complicating the interpretation of gravitational lensing data. Such blending can occur in binary star systems or through the coincidental alignment of stars. The depth of astronomical surveys, like the LSST, means a significant number of galaxies observed will have overlapping images with others. For instance, the Hyper Suprime-Cam (HSC) Wide survey found that 58\% of galaxies share their luminous pixel regions with other objects \cite{bosch2018hyper}, while LSST predicts a 62\% overlap  \cite{sanchez2021effects}.

The Zwicky Transient Facility (ZTF) stands out as an expansive sky survey designed to capture a broad section of the sky simultaneously, albeit with a trade-off between wide-field coverage and resolution. Prioritizing the ability to survey extensive areas, ZTF sacrifices resolution to achieve its primary objective: the timely detection of transients such as supernovae, gamma-ray bursts, variable stars, asteroids, and exoplanets. While ZTF may not boast the precision desired for certain scientific endeavors, its vast field-of-view results in the accumulation of substantial volumes of data. Consequently, this abundance of data presents an intriguing opportunity for scientists to explore the potential of machine learning methodologies to extract meaningful insights.

A plethora of literature exists on machine learning models that leverage color and morphological information as inputs for neural networks, and ensemble methods like random forests \cite{ odewahn1992automated, kim2015hybrid}. However, the challenge lies in generalizing data features from high-dimensional datasets without relying extensively on expert knowledge. 

One traditional approach is to manually engineer features as inputs to classification models. \cite{Sevilla2015} used Random Forests and \cite{odewahn1992automated} applied deep neural networks (DNN) to successfully distinguish stars from galaxies. \cite{Kim_2016} applied convolutional neural networks (CNN) to SDSS data, 5 channels and 44$\times$44 pixels. These models have been noted to incorrectly classify out-of-distribution samples with confidence \cite{scheirer2012toward,yang2021generalized}. \cite{muyskens2021MuyGPs} present a computationally efficient Gaussian process (GP) algorithm that has been applied successfully to other similar astronomical applications, e.g. star-galaxy disambiguation and galaxy blend detection. MuyGPs uses an approximate k-nearest neighbors with $k$ neighbors in batches of size $b$ to achieve a computational complexity for model fitting of $O({bk}^3)$ compared to a maximum likelihood expectation method that is $O(n^4)$, where $n$ is the training set size. In the MuyGPs framework, a threshold can be defined based on the model's prediction confidence, and the ambiguous samples can be further investigated. We compare all these models (Random Forests, DNN, CNN, GP) using data with different normalizations and feature embeddings.

GP models are widely utilized in nonlinear modeling \cite{gelfand2010handbook, stein1999interpolation, muyskens2021MuyGPs}, wherein the process is considered a GP if any finite set of $n$ realizations follows a multivariate normal distribution. Conventionally, it is assumed that the covariance matrix depends on a pairwise kernel function, which in turn relies on input hyperparameters. However, the challenge lies in the learning of these hyperparameters (denoted as $\theta$), a task that is often difficult and computationally expensive. The computation and storage of the covariance matrix incur a cost of $O(n^2)$, while the realization of Gaussian process regression (GPR) and the evaluation of likelihood, essential components in the conventional training of $\theta$, have a higher cost of $O(n^3)$. Consequently, training a complete GP model becomes prohibitively expensive for large $n$ when utilizing common hardware.

We introduce a novel and cost-effective Gaussian Process (GP) model to analyze a collection of images from the Zwicky Transient Facility (ZTF). The analysis categorizes ZTF images into three distinct types: a single star, a binary star system (blended star), or a blend of two or more stars due to their alignment along the same line-of-sight (LOS). Typically, differentiating these situations demands resource-intensive techniques like spectroscopy, acquiring detailed images from space telescopes, or securing higher-resolution images with ground-based observatories. The GP model we propose offers a scalable solution to process ZTF's extensive datasets efficiently, sidestepping the high computational costs traditionally associated with this task. By employing specific normalization and data embedding techniques before utilizing machine learning algorithms, the GP model achieved an impressive 83.8\% accuracy, outperforming random forests, the next best model.

\section{Data} \label{data}
\begin{figure}[h!]
    \centering
    \includegraphics[scale=0.5]{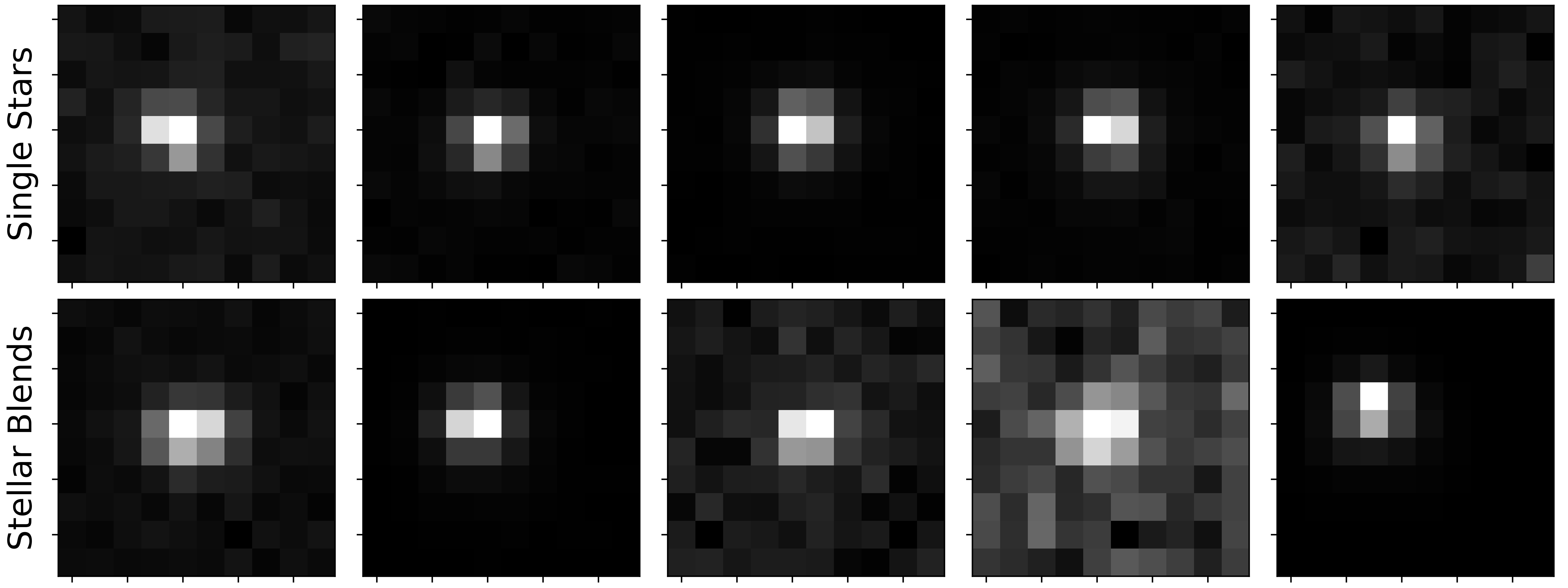}
    \caption{Example images of single stars and stellar blends}
    \label{images}
\end{figure}

The ZTF image dataset comprises 27,253 one-channel 10x10 pixel images with binary classification labels. Specifically, there are 15,110 images of single stars and 12,143 of stellar blends. The stellar blends further break down into two classes: 7,414 blended stars and 4,729 binary stars. Thus, this dataset presents both binary and multi-class classification challenges. Notably, visually distinguishing these three classes is challenging for the human eye, underscoring the suitability of an algorithmic approach capable of identifying patterns that might elude even a trained observer. All images in the dataset are ZTF i-band images, corresponding to the 700-900nm range in telescope filters. Consequently, each image captures emissions solely from the object(s) within this wavelength range, providing a focused perspective on astronomical features in this specific spectral domain.
Figure \ref{images} shows plots of samples of single stars and blended stars.

To delineate between single stars and stellar blends (non-binary), selections were made using a subset of data from the DESI Legacy Imaging Survey \cite{dey2019overview}. Criteria for these selections included various catalog values, such as checking the distance between any two stars in the catalog. To ensure visibility in ZTF images, additional cuts were made based on magnitudes corresponding to limits defined by the ZTF survey. The binary star class was determined through a binary catalog. Given the inherent complexities of astronomical cataloging, potential errors in labeling binaries, and other challenges, there is a chance that a small number of objects may be mislabeled. However, these instances are expected to have minimal impact on the overall outcome of the classification task. Challenges such as cutouts inadvertently excluding one of the stars in a blend or the presence of a nearby bright star not part of the blend are acknowledged and retained within the dataset.

\section{Methods}
\subsection{Normalization and Embedding} \label{data-norm}
Effective data normalization is crucial for enhancing machine learning model performance, especially for datasets like ZTF images. These images, despite their low resolution, contain features at various scales. Thus, it is challenging to separate them during model training.

\begin{figure}[h!]
    \centering
    \includegraphics[scale=.5]{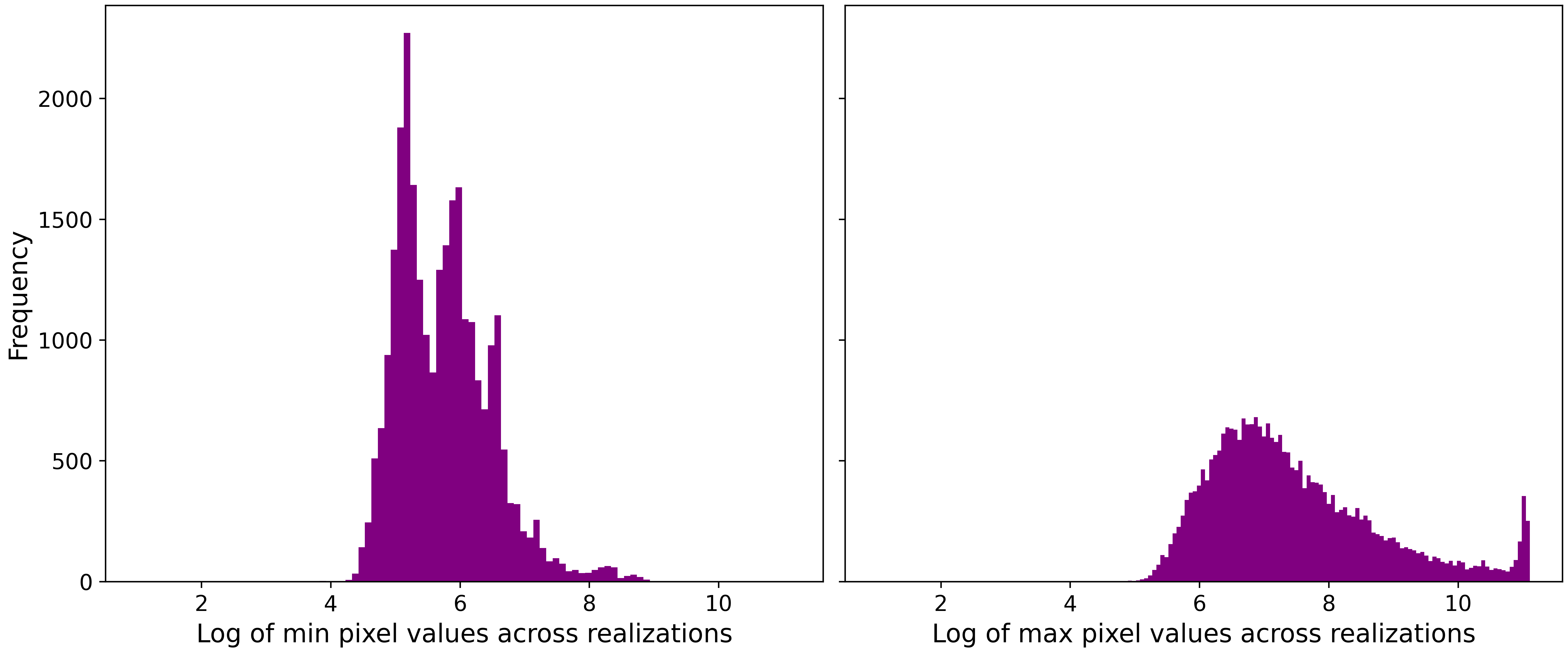}
    \caption{Distributions of log min and max pixels across different realizations of the ZTF images}
    \label{pixel-distributions}
\end{figure}

This work explores the application of several normalization techniques on ZTF images, including log, min-max, and $r^\text{th}$ root transformations. The selection of these techniques is motivated by the specific distribution of pixel values within the ZTF dataset (refer to Figure \ref{pixel-distributions} for details). Each method offers advantages:
1) log normalization caters to the observed distribution of pixel intensities.
2) Min-max normalization leverages the spread of minimum and maximum pixel values across the entire dataset, promoting numerical stability and efficient computations.

These normalization techniques address the inherent challenges of separating features in low-resolution, multi-scale ZTF images. By enhancing feature separability, we aim to significantly improve the effectiveness of machine learning models in extracting meaningful information from this data.

Building upon the log function's effectiveness, we introduce a novel class of normalizations based on the $r^\text{th}$ root transformation (where $r$ is a value between 0 and 1).  These transformations aim to enhance feature separability within ZTF images.

For mathematical clarity, we express the normalizations as follows:
\begin{align}
\log(\bm{x}_i) &= \log |\bm{x}_i|, \label{log}\\
\text{root}_r(\bm{x}_i) &= |\bm{x}_i|^r. \label{rroot}
\end{align}
In both equations, the absolute value ensures we only consider the magnitude of pixel intensities, and all operations are applied to each pixel value within the image $\bm{x}_i$ independently.

This approach leverages the strengths of both log and power-law transformations, potentially leading to improved feature separation in ZTF images compared to traditional methods.

To further enhance the proposed normalizations, we combine them with min-max scaling strategies. These strategies introduce two variants: local and global min-max scaling.
Local min-max scaling computes the minimum and maximum values for each ZTF image individually and uses them to scale pixel intensities within that image.
In contrast, global min-max scaling calculates the minimum and maximum values across all pixels in the entire training dataset. These values are then used to scale pixel intensities in every image. 
Given the $i^\text{th}$ image as $\bm{x}_i$, let  $\bm{x}_i^{kl}$ be the pixel at the $k$-$l$ entry. Then mathematically,
\begin{align}\label{loc-scaler}
\text{loc}_{minmax}(\bm{x}_i) &= \frac{g(\bm{x}_i) - \underset{k,l}{\min} \, g(\bm{x}_i^{kl})}{\underset{k,l}{\max} \, g(\bm{x}_i^{kl}) - \underset{k,l}{\min} \, g(\bm{x}_i^{kl})},\\
\text{glob}_{minmax}(\bm{x}_i) &= \frac{g(\bm{x}_i) - \underset{k,l}{\min} \, g(\bm{x}_i^{kl})}{\underset{j,k,l}{\max} \, g(\bm{x}_j^{kl}) - \underset{j,k,l}{\min} \, g(\bm{x}_j^{kl})},
\label{glob-scaler}
\end{align}
where the function $g$ applied to $\bm{x}_i$, subtraction and division are  component-wise operations. 
 
The key difference lies in how minimum and maximum values are determined.
The choice between local and global scaling depends on the distribution of minimum and maximum pixel values within the ZTF dataset (see Section~\ref{data} for details). We expect their performance to vary depending on the specific data characteristics.

The proposed $r^\text{th}$ root normalization offers an additional benefit. The parameter $r$ can be learned during the training process. This allows the model to automatically identify a stable value within the range (0, 1) that best captures the inherent distribution of pixel intensities. This not only improves computational efficiency but also potentially facilitates the extraction of meaningful features by the machine learning model.

\begin{figure}[h!]
\includegraphics[scale=.4]{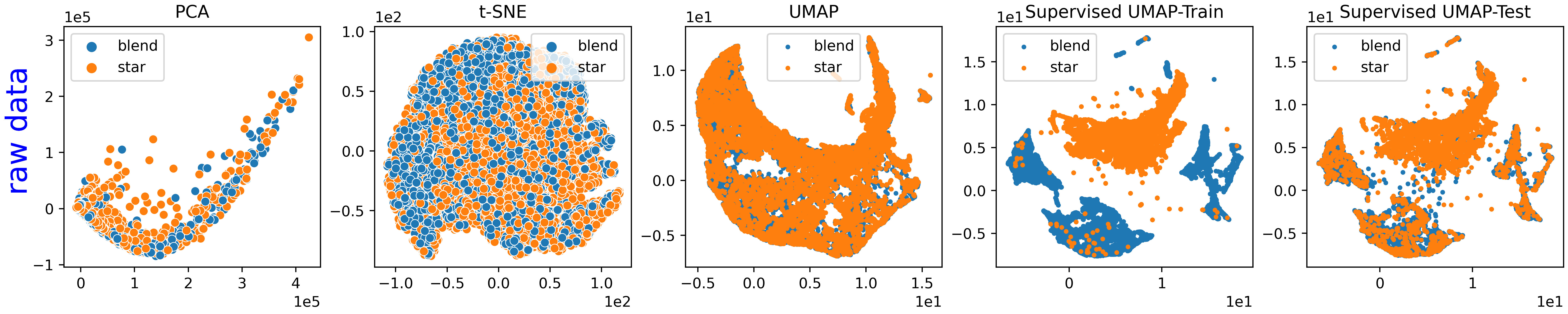}
\vspace{0.1cm}
\includegraphics[scale=.4]{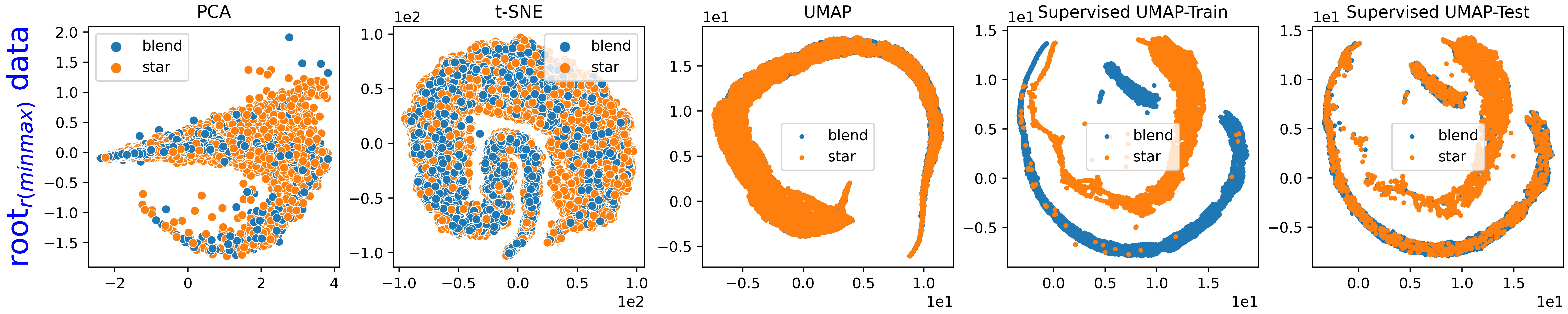}
\vspace{0.1cm}
\includegraphics[scale=.4]{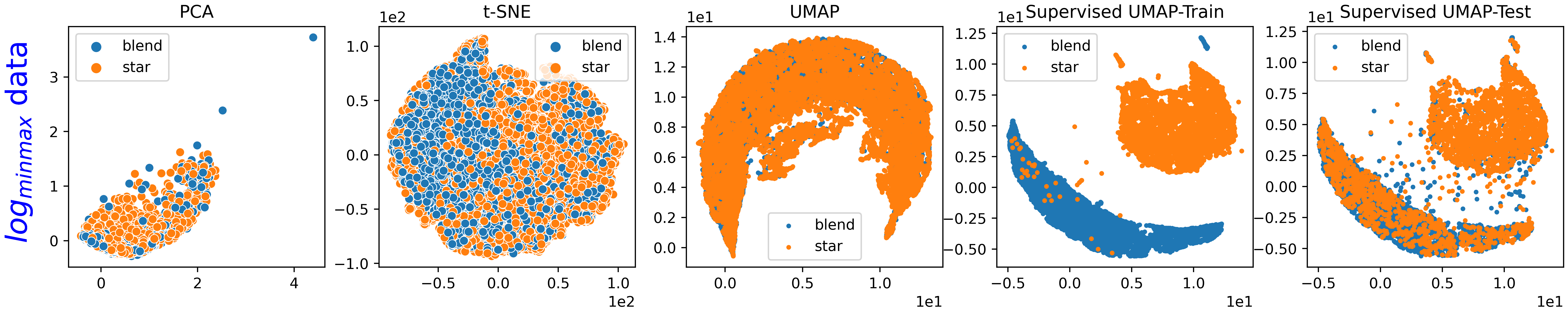} 
\vspace{0.1cm}
\caption{Data Normalizations and Embedding}\label{embedding}
\label{raw-embedding}
\end{figure}

We use various techniques to visualize and embed the normalized ZTF image data. 
Principal Component Analysis (PCA) of \cite{hotelling1933analysis} is a well-established method for dimensionality reduction and visualization, while t-distributed Stochastic Neighbor Embedding (t-SNE) and Uniform Manifold Approximation and Projection (UMAP) are more recent techniques known for their effectiveness in visualizing complex, high-dimensional data \cite{van2008visualizing, McInnes_2018}.
Figure~\ref{embedding} showcases the embedding results for the raw and normalized data using PCA, t-SNE, and UMAP. This visualization allows us to compare the effectiveness of different normalization techniques in enhancing the separability of features within the ZTF images.

While various normalization techniques were applied, unsupervised UMAP embeddings (similar to PCA and t-SNE) exhibited limited data separability. This suggests that the inherent structure of the data might not be fully captured by these unsupervised methods.
In contrast, supervised UMAP, which leverages class labels during embedding, yielded a significant improvement in separability, particularly for the testing dataset. Notably, log and $r^\text{th}$ root normalizations combined with min-max scaling appeared most effective.

Building upon these findings, we plan to incorporate the results of supervised UMAP into our machine learning models. This approach will allow us to analyze the dimensions of the extracted feature space and explore its effectiveness in distinguishing between individual sources and stellar blends within the ZTF data.

\subsection{Gaussian Process Classification}
Gaussian process (GP)  models provide a Bayesian non-parametric method for addressing supervised machine learning tasks in both regression and classification \cite{challis2015gaussian}. Here, we focus on Gaussian process classification (GPC). For an in-depth overview on Gaussian process regression and classication, see \cite{rasmussen2006gaussian}.

Consider the dataset $\{(\bm{x}_i, y_i )\}_{i=1}^n$ of feature points $\bm{x}_i$ from a domain $\mathcal{X}$ associated with binary class labels $y_i \in \{-1, +1\}$. We aim to predict the class membership for an unseen test point $\bm{x}^*$ \cite{kim2006bayesian, nickisch2008approximations, kim2006bayesian}. 
We assume that the likelihood of class labels, given the input $\bm{x}_i$, depends on the value of an underlying real-valued function,  $f : \mathcal{X} \to \mathbb{R}$. Specifically, in binary classification, once $f(\bm{x}_i)$ is determined, the probability of a class label becomes independent of all other factors. This is realized by employing the latent function $f$, whose output is transformed into the unit interval using one of sigmoid, probit, heavyside functions, etc, as summarized in equation \eqref{activations}.
\begin{equation}\label{activations}
p(y_i \big| \theta, \bm{x}_i) = 
\begin{dcases}
\begin{aligned}
H(y_i f(\bm{x}_i))  && \text{sign function}\\
erf(y_if(\bm{x}_i))  && \text{probit function}\\
\frac{1}{1 + exp(-y_if(\bm{x}_i))}  && \text{sigmoid function}  
\end{aligned}
\end{dcases}
\end{equation}
where $H(\eta) = 1$ if $\eta > 0$ and $0$ otherwise,
$erf(\eta) = \int_{-\infty}^\eta \frac{1}{\sqrt{2\pi}} e^{-\frac{x^2}{2}} \, dx$ \cite{kim2006bayesian}. 

In practice,  any of the activation functions in equation \eqref{activations} can be modulated into the sign function
\begin{equation}\label{MuyGPs-activation}
\text{sgn}(f(\bm{x}_i)) =  
\begin{dcases}
\begin{aligned}
+1 && f(\bm{x}_i) > 0\\
-1 && f(\bm{x}_i) < 0,
\end{aligned}
\end{dcases}
\end{equation}
with $0$ as the decision boundary.
We then get the benefit of treating GPC as a regression problem with the labels $\{-1, +1\}$ as regression targets \cite{muyskens2022star, rasmussen2006gaussian}.

Given any finite set of inputs $X = \{\bm{x_1}, \cdots , \bm{x_n}\} \subset \mathcal{X}$ with a corresponding vector of labels $\bm{y}$,
the Bayesian model representing the Gaussian Process (GP) prior on the function $f$ becomes \cite{goumiri2020star}:
\begin{align}
\frac{\bm{y}}{\sigma} &= \bm{f} + \varepsilon \\
\bm{f} &= [ f(\bm{x}_1), \cdots , f(\bm{x}_n)]^T \sim \mathcal{N} (0, K(X,X)) \label{pd-data}\\
\varepsilon &\sim \mathcal{N} (0, \tau^2I_n) \label{error}
\end{align}
where $\bm{y}$ is an $n$-vector representing evaluations of $f$ on elements of $X$. 
$K(X,X)$ is an $n \times n$ positive semi-definite covariance matrix with elements $(i, j)$ given by $k(\bm{x}_i, \bm{x}_j) = \text{cov}(f(\bm{x}_i), f(\bm{x}_j))$. Note that $\varepsilon$ represents homoscedastic Gaussian noise affecting the measurement of $f$, and $\tau^2$, the variance of the unbiased homoscedastic noise. 
The covariance matrices implicitly depend on a learnable parameter, $\theta$, and 
$\sigma$ is a scale factor proportional to the overall variance in the prior \cite{buchanan2022gaussian}. 

Now, consider $m$ unlabeled embedded images $X^* = \{\bm{x^*_1}, \cdots, \bm{x^*_m}\} \subset \mathcal{X}$. The joint distribution of $\bm{y}$ and the response $\bm{f^*}$ of $f$ applied to $X^*$ is given by:
\begin{align}\label{predict}
\begin{bmatrix}
\bm{y} \\
\bm{f}^*
\end{bmatrix}
&= \mathcal{N}\left(0, 
\sigma^2 
\begin{bmatrix}
K(X,X) + \tau^2 I_n & K(X,X^*) \\
K(X,X^*) & K(X^*, X^*)
\end{bmatrix}\right)
\end{align}

Here, $K(X^*,X) = K(X, X^*)^T$ is the cross-covariance matrix between $X^*$ and $X$, where the $(i, j)$th element of $K(X^*,X)$ is $k(\bm{x}_i^*, \bm{x}_j)$. Meanwhile, $K(X^*,X^*)$ has $(i,j)$th element
$k(\bm{x}_i^*, \bm{x}_j^*)$. Applying Bayes' rule to \eqref{pd-data} and \eqref{predict}, we can compute the posterior distribution of the response $\bm{f^*}$ on $X^*$ in closed form as in \cite{muyskens2022star}:
\begin{align}\label{posterior}
\bm{f}^* \big| (X^*, X,  \bm{y}) &\sim \mathcal{N}(\hat{\bm{f}}^*, \sigma^2 C) \nonumber\\
     \hat{\bm{f}}^*        & = K(X^*,X) (K(X,X) + \tau^2 I_n)^{-1} \bm{y} \nonumber\\
                         C  & =  K(X^*, X^*) - K(X^*,X) (K(X,X) + \tau^2 I_n)^{-1} K(X,X^*)
\end{align}
We employ a straightforward discriminator based on the posterior of the response on $X^*$: if the model output  $\hat{\bm{f}}^*_i \ge 0$ then $\bm{x}_i^*$ is classified as a single star; otherwise, it is categorized as a stellar blend. Importantly, we can assess the variance of this prediction by inspecting the $i$th diagonal entry of the matrix $C$. This enables us to automatically identify situations where the posterior variance is substantial, indicating that the model lacks confidence in its prediction \cite{goumiri2020star}.

Conventional techniques for training Gaussian Process (GP) models estimate covariance parameters via grid search cross-validation, Bayesian analysis with Markov Chain Monte Carlo (MCMC), or maximum likelihood estimation using the log-likelihood. For large datasets, however, these methods are computationally demanding, requiring at least $O(n^3)$ computation and $O(n^2)$ memory \cite{muyskens2021MuyGPs}. 

MuyGPS addresses this challenge by 
employing a local kriging approach, focusing on the $k$ nearest neighbors of each prediction point instead of the entire dataset. This method reduces the computational complexity to $O(nk^3)$, making it linear in the sample size $n$ and far more tractable, more so when $k \ll n$, see  \cite{muyskens2021MuyGPs} for further details.
Specifically, let $X_{N_i}$ be the set of training observations nearest to $\bm{x}_i$, and $y_{N_i}$  their corresponding labels. The prediction at $\bm{x}_i$ is given by:
$$
\hat{f}_i^{NN} = K(\bm{x}_i, X_{N_i}) K(X_{N_i}, X_{N_i})^{-1} y_{N_i}
$$
where $K(\bm{x}_i, X_{N_i})$ is the cross-covariance between $\bm{x}_i$ and its nearest neighbors, and $K(X_{N_i}, X_{N_i})$ is the covariance among these neighbors. The corresponding posterior variance at $\bm{x}_i$ is:
$$
\text{Var}(\bm{x}_i|X_{N_i}) = K(\bm{x}_i, \bm{x}_i) - K(\bm{x}_i, X_{N_i}) K(X_{N_i}, X_{N_i})^{-1} K(X_{N_i}, \bm{x}_i)
$$
MuyGPs uses batched leave-one-out cross-validation to optimize hyperparameters of the covariance kernel without requiring potentially costly likelihood computations. It builds on earlier techniques that make use of the closest neighborhood structure of the data. This approach ensures computational efficiency while still maximizing the predictive and uncertainty quantification power of Gaussian processes \cite{muyskens2021MuyGPs}.

The posterior distribution, as described in equation \eqref{posterior} is contingent on the choice of the kernel function. We specifically opt for the Matérn kernel, a stationary and isotropic kernel commonly utilized in spatial statistics GP literature for its flexibility and favorable properties \cite{stein1999interpolation}. Despite the option of using the radial basis function (RBF) kernel, we find that applying either the Matérn or RBF kernel to the ZTF image data does not result in a substantial effect. This observation leads us to select the Matérn kernel, emphasizing its capacity to provide enhanced control over the smoothness of the GP compared to the RBF kernel. The general expression for the kernel is 
\begin{align}
k_{\text{matern}} (\bm{x}, \bm{x}') 
=
\sigma^2 \frac{2^{1-\nu}}{\Gamma(\nu)} 
\biggl( \sqrt{2\nu} \frac{\Vert \bm{x}- \bm{x}'\Vert^2_2}{\ell}  \biggr)^\nu K_\nu 
\biggl(\sqrt{2\nu}\frac{\Vert\bm{x}-\bm{x}' \Vert^2_2}{\ell}  \biggr),
\end{align}
where $\nu > 0$ is smoothness parameter, $\ell > 0$, a correlation length scale hyperparameter, $\sigma^2 > 0$, a scale parameter, $\Gamma$, the Gamma function, and $K_\nu (\cdot)$, the modified Bessel function of the second kind. 

A comprehensive explanation of the estimation process for all the hyperparameters contained within $\theta$ can be found in \cite{muyskens2022star}, as well as in the MuyGPs package \cite{MuyGPs2021git}. For the sake of brevity, we exclude those details here.

\section{Results}
Our analysis compares the effectiveness of a novel machine learning model (MuyGPs) against existing ones in classifying between single sources and stellar blends within ZTF astronomical images. Table \ref{results-1} explores the performance under various normalization techniques applied to the low-resolution data.

\begin{table}[ht] \label{results-1}
\begin{tabular}{lcc}
\hline
\textbf{Model}           & \textbf{Normalization}                              & \textbf{Accuracy (\%)}                 \\ \hline
Random   Forests & Raw Pixel                                                         & 80.8\\
DNN             & Raw Pixel                                                         & 74.8\\
CNN             & Raw Pixel                                                         & 62.8\\
\textbf{MuyGPs}          & \textbf{Raw Pixel}                                                          & \textbf{80.9}\\ 
\hline
Random   Forests & log Local Min-Max                                                       & 82.3\\
DNN             & log Local Min                                                       & 73.4\\
CNN             & log Local Min                                                           & 74.9\\
\textbf{MuyGPs}          & \textbf{ log Local Min-Max}                                    & \textbf{83.4}\\  
\hline
Random   Forests &   log Global Min-Max                                                     & 82.3\\
DNN             & log Global Min                                                       & 74.7\\
CNN             &   log Global Min-Max                                                     & 74.4\\
\textbf{MuyGPs}          & \textbf{ log Global Min}                                 &\textbf{82.4}\\  
\hline
Random   Forests & $r^{\text{th}}$ root  Local Min-Max                               & 83.1\\
DNN             & $r^{\text{th}}$ root Local Min-Max                                & 76.8\\
CNN             & $r^{\text{th}}$ root  Local Min-Max                               & 74.3\\
\textbf{MuyGPs} & \textbf{$r^{\text{th}}$ root  Local Min-Max}                      & \textbf{83.8}\\ 
\hline
Random   Forests & $r^{\text{th}}$ root  Global Min-Max                              & 80.6\\
DNN             & $r^{\text{th}}$ root Global Min-Max                               & 77.7\\
CNN             & $r^{\text{th}}$ root  Global Min-Max                              & 79.0\\
\textbf{MuyGPs} & \textbf{$r^{\text{th}}$ root  Global Min-Max}                     & \textbf{82.1}\\ 
\hline
\end{tabular}
\caption{Comparison of different model performances.}
\label{tab:model_benchmark}
\end{table}

Our experiments (presented in Table~\ref{results-1}) utilized a common data split: 80\% for training and 20\% for testing. We employed two main types of machine learning models for comparison: random forests, a well-established model and we implemented various deep neural networks (DNN) architectures (details in Appendix, Table~\ref{dnn-model}).

Interestingly, while Convolutional Neural Networks (CNNs) are typically the go-to approach for image data, their performance on our low-resolution ZTF dataset was competitive with other baselines, especially considering the limited training data. Table~\ref{cnn-model} (Appendix) summarizes the CNN architecture. We found that limiting the number of convolutional layers was crucial to prevent losing information from the small (10x10) images. Notably, the absence of additional fully connected layers hindered performance, suggesting that these layers play a significant role in the CNN's effectiveness for this specific data.
For our novel machine learning model, MuyGPs, we used the following parameters: nearest neighbors (30), batch size (200), cross-entropy loss function, Matern kernel, and F2 metric.

\begin{figure}[h!]
    \centering
    \includegraphics[scale=0.7]{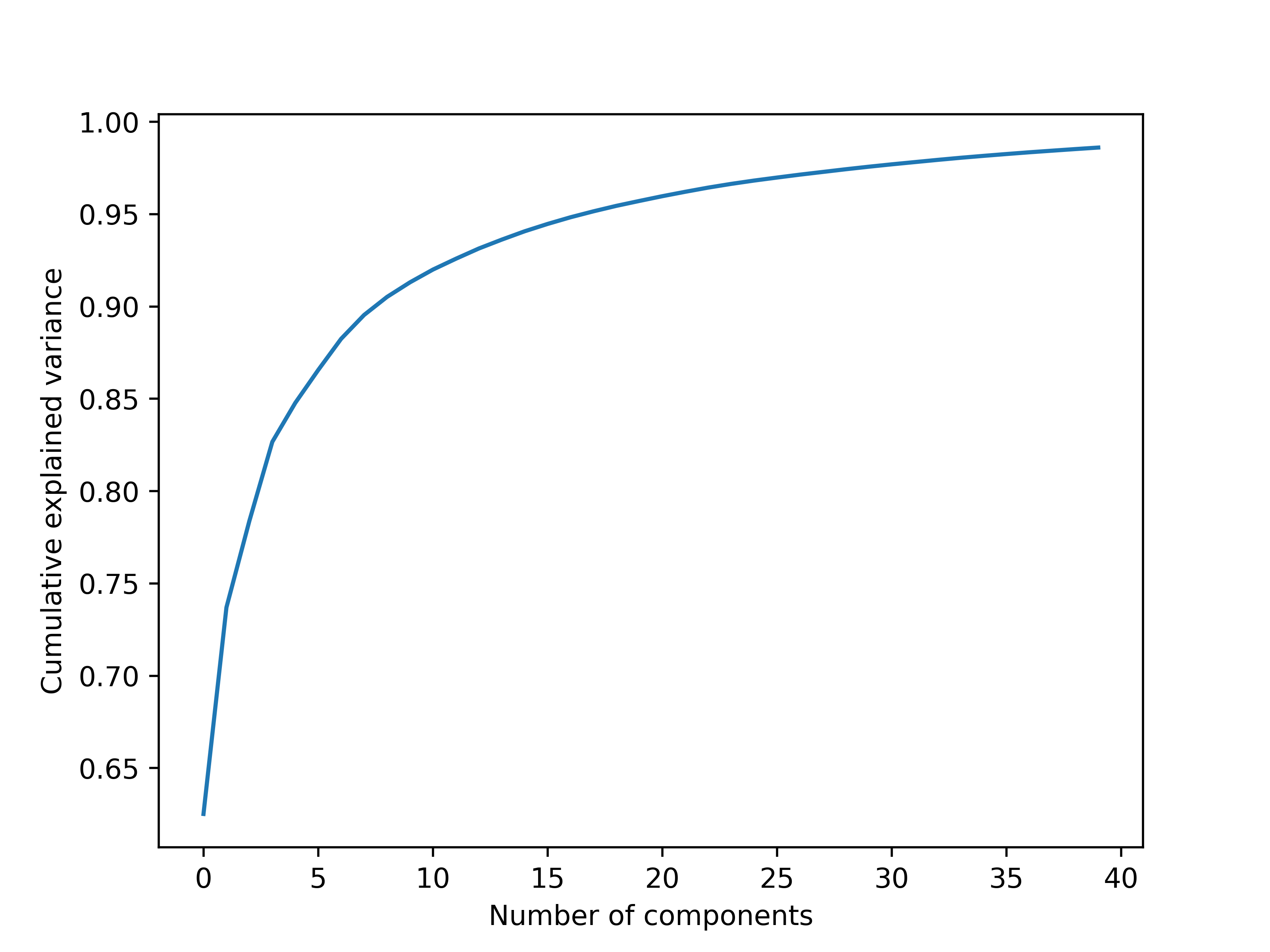}
    \caption{PCA Scree Plot for Raw Image Data}
    \label{scree-plot}
\end{figure}

We investigate the effects of data embedding on the performance of our proposed model, MuyGPs. Here, we focus on models achieving the highest accuracy for each normalization technique (details in Table~\ref{results-1}, Appendix).
Since baseline models like DNNs and CNNs inherently involve dimensionality reduction, we didn't expect significant performance improvement with embedded data. Conversely, we explored embedding techniques with MuyGPs in hopes of enhancing its performance.
To determine the appropriate dimensionality for the embedded data, we analyzed the raw image data using Principal Component Analysis (PCA) and visualized the results in a scree plot (Figure~\ref{scree-plot}). Scree plots for normalized data exhibited similar patterns and are omitted for brevity.

Our analysis of the raw image data using Principal Component Analysis (PCA) revealed that a relatively low number of principal components (around 5 to 10) can explain a significant portion (up to 90\%) of the total variation (visualized in the scree plot, Figure~\ref{scree-plot}). This suggests that the underlying structure of the data resides in a low-dimensional space.
Building upon this finding, we explored feeding the outputs of PCA embedding for different data normalizations into our MuyGPs model. The results of these experiments are summarized in Figure~\ref{MuyGPs-pca-embedding} (Appendix).

Our experiments employing PCA embedding for various normalization techniques with MuyGPs yielding valuable insights (see Figure~\ref{MuyGPs-pca-embedding}). We notice that
combining min-max scaling with the $r^\text{th}$ root normalization generally led to better performance compared to using the $r^\text{th}$ root normalization alone. This is evident in the figure, where the min-max region exhibits higher accuracy.
The $r^\text{th}$ root normalization without min-max scaling required a larger number of principal components to achieve similar performance, suggesting that min-max scaling improves the data's separability for MuyGPs.
An interesting trend emerges regarding the choice of the $r$ parameter. When using min-max scaling, accuracy increased as the $r$ value increased. However, the opposite was true for the $r^\text{th}$ root normalization without min-max scaling.
Among all tested configurations, the best performing model achieved an accuracy of 82.315\%. This model combined log normalization with min-max scaling and utilized 34 principal components from the PCA embedding. The second-best model employed the $r^\text{th}$ root normalization with min-max scaling at r=0.7931, achieving an accuracy of 81.82\% with 26 principal components.

While PCA embedding exhibits varying performance for MuyGPs depending on the number of principal components, supervised UMAP embedding yielded more consistent results (findings visualized in Figure~\ref{MuyGPs-umap-embedding}). We observe that the choice of the number of UMAP components had minimal impact on MuyGPs performance. High accuracy was achieved even with as few as 4 components, which suggests a more efficient capture of the relevant data structure by supervised UMAP. Similar to PCA embedding, the combination of log normalization and min-max scaling (norm-21) remained the top performer. However, the performance difference compared to other normalizations was less significant for supervised UMAP.
Interestingly, the $r^\text{th}$ root normalization achieved a very close accuracy (81.875\%) using only 4 UMAP components, compared to the 81.893\% achieved by norm-21 with 31 components.
Notably, for both local and global min-max scaling scenarios, the best performing model in terms of accuracy remained norm-21, with results similar to those observed for local min-max with PCA embedding.

\begin{figure}[h!]
    \centering
    \includegraphics[scale=0.3]{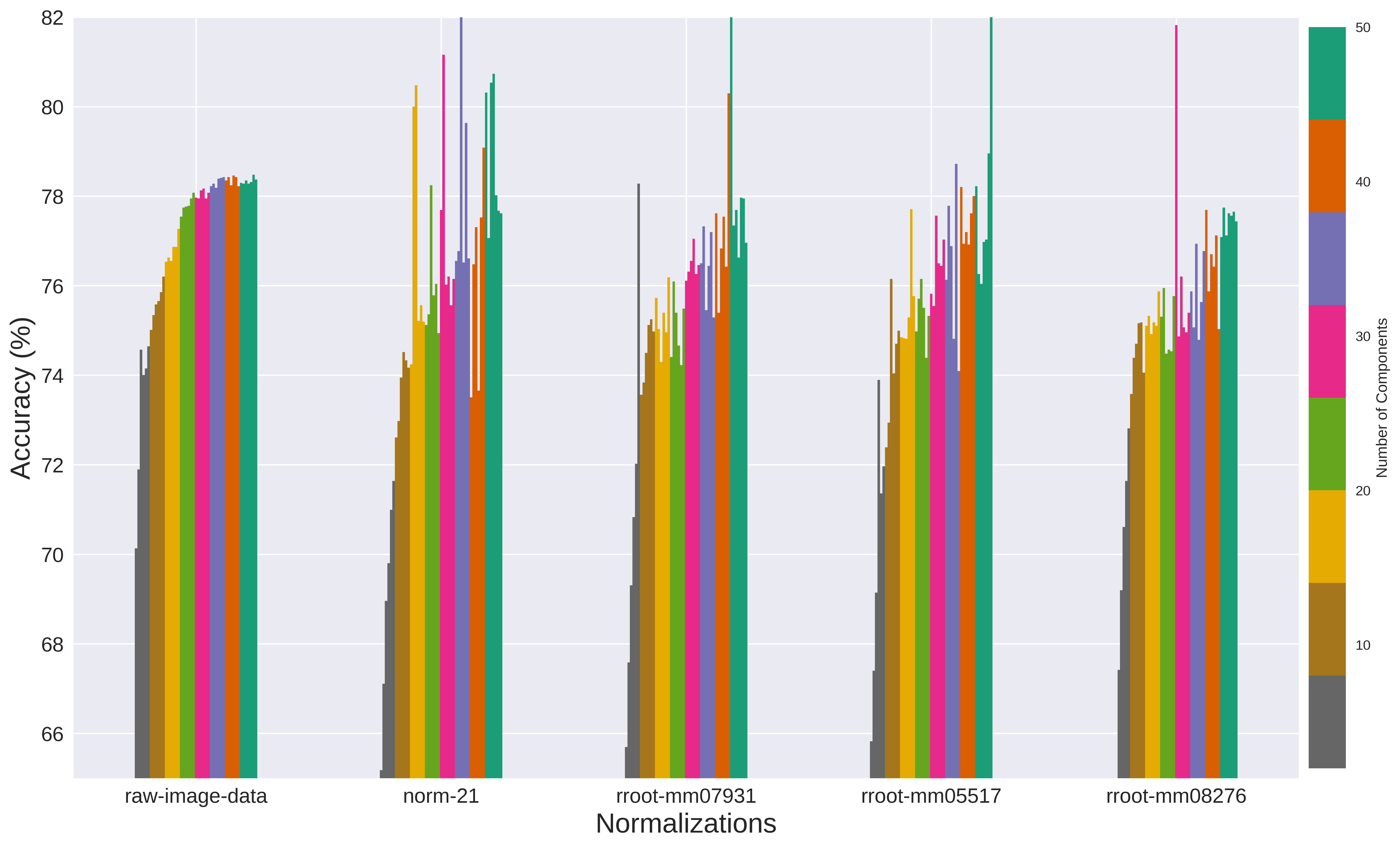}
    \caption{MuyGPs top 4 normalizations with PCA embedding across 20-50 components, compared to the raw data }
    \label{MuyGPs-top-pca-embedding} 
\end{figure}

\begin{figure}[h!]
    \centering
    \includegraphics[scale=0.3]{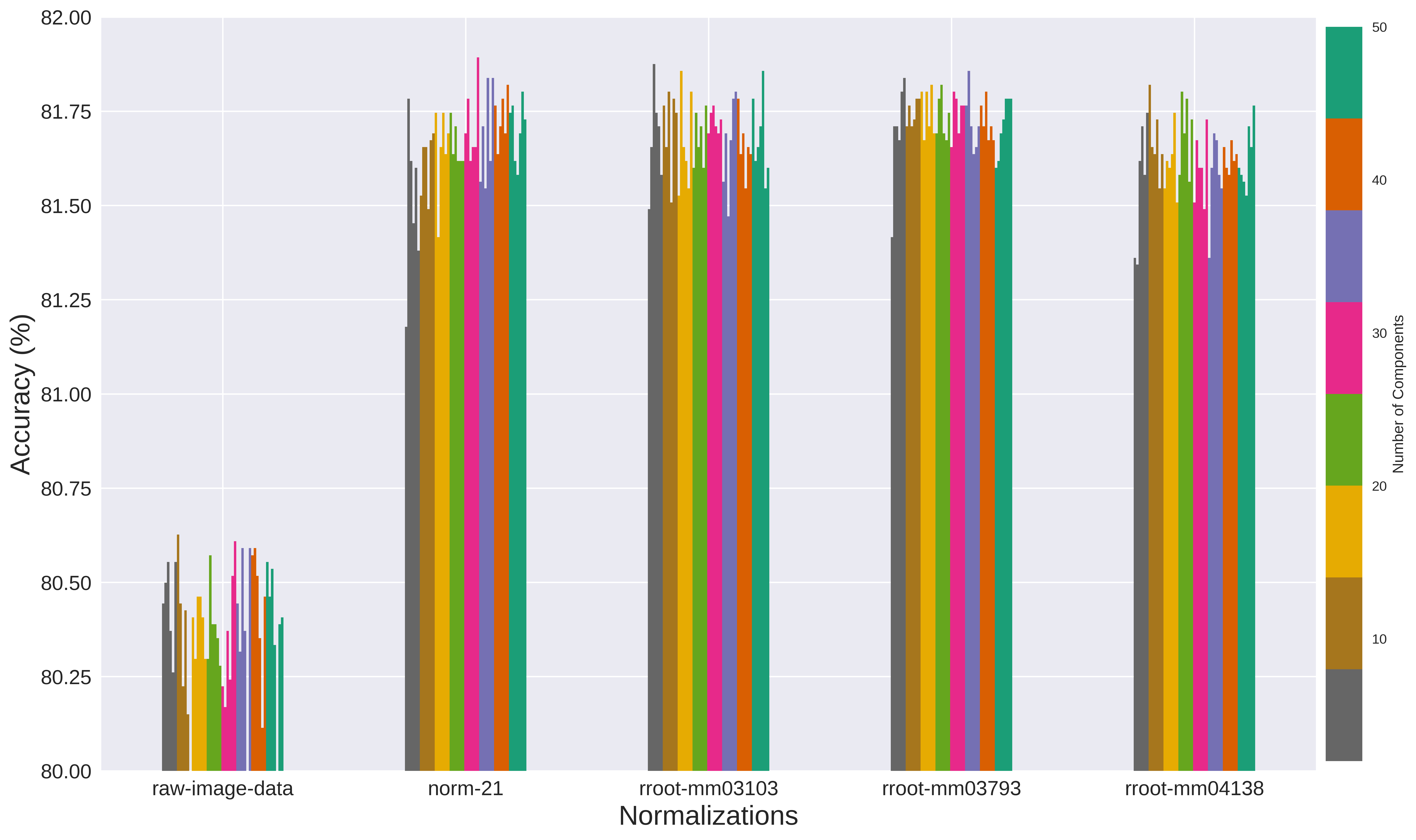}
    \caption{MuyGPs top 4 normalizations with supervised UMAP embedding across 20-50 components }
    \label{MuyGPs-top-umap-embedding} 
\end{figure}

The results presented in Figure~\ref{MuyGPs-top-umap-embedding} reveal a significant advantage of supervised UMAP embedding. MuyGPs performance remains relatively constant regardless of the number of UMAP components used. This suggests that supervised UMAP efficiently captures the essential structure of the data in a lower-dimensional space, offering significant computational benefits.
The bar charts in Figures~\ref{MuyGPs-top-umap-embedding} and \ref{MuyGPs-top-pca-embedding} further emphasize the importance of data normalization. Different normalization techniques clearly lead to varying performance levels for MuyGPs, even with supervised UMAP. This highlights the crucial role of proper data preparation in achieving optimal model performance.

\begin{figure}[h!]
    \centering
    \includegraphics[scale=0.6]{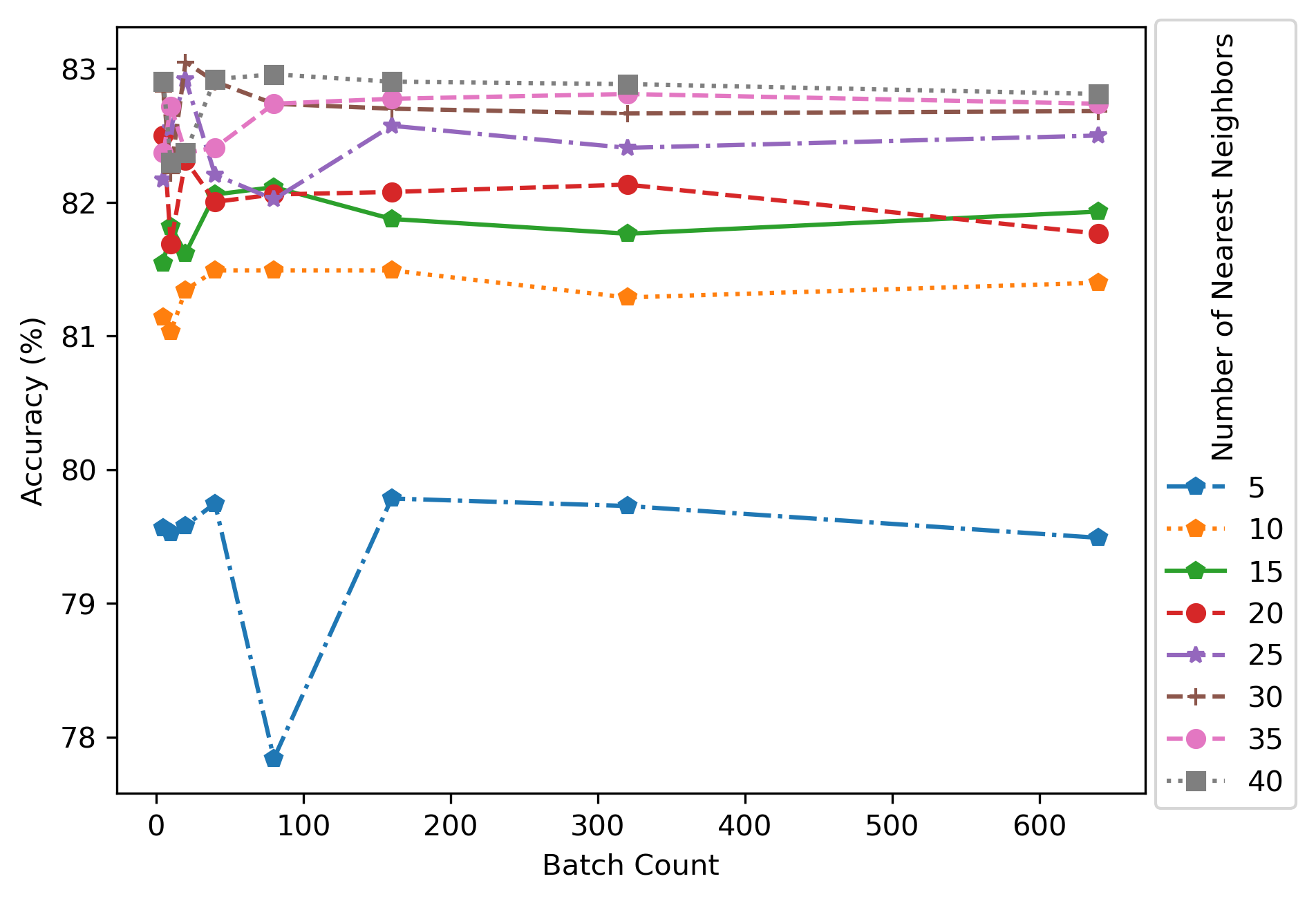}
    \caption{MuyGPs Accuracies for Different Batch and NN Counts }
    \label{MuyGPs-batch-nn-counts} 
\end{figure}

We further investigate the influence of various hyperparameters on MuyGPs accuracy. One key finding, as shown in Figure~\ref{MuyGPs-batch-nn-counts}, is that MuyGPs performs well with smaller batch sizes. Interestingly, increasing the batch size does not lead to improved accuracy. This suggests that MuyGPs can be computationally efficient compared to full GP classification, which typically benefits from larger batch sizes.

\begin{figure}[h!]
    \centering
    \includegraphics[scale=0.6]{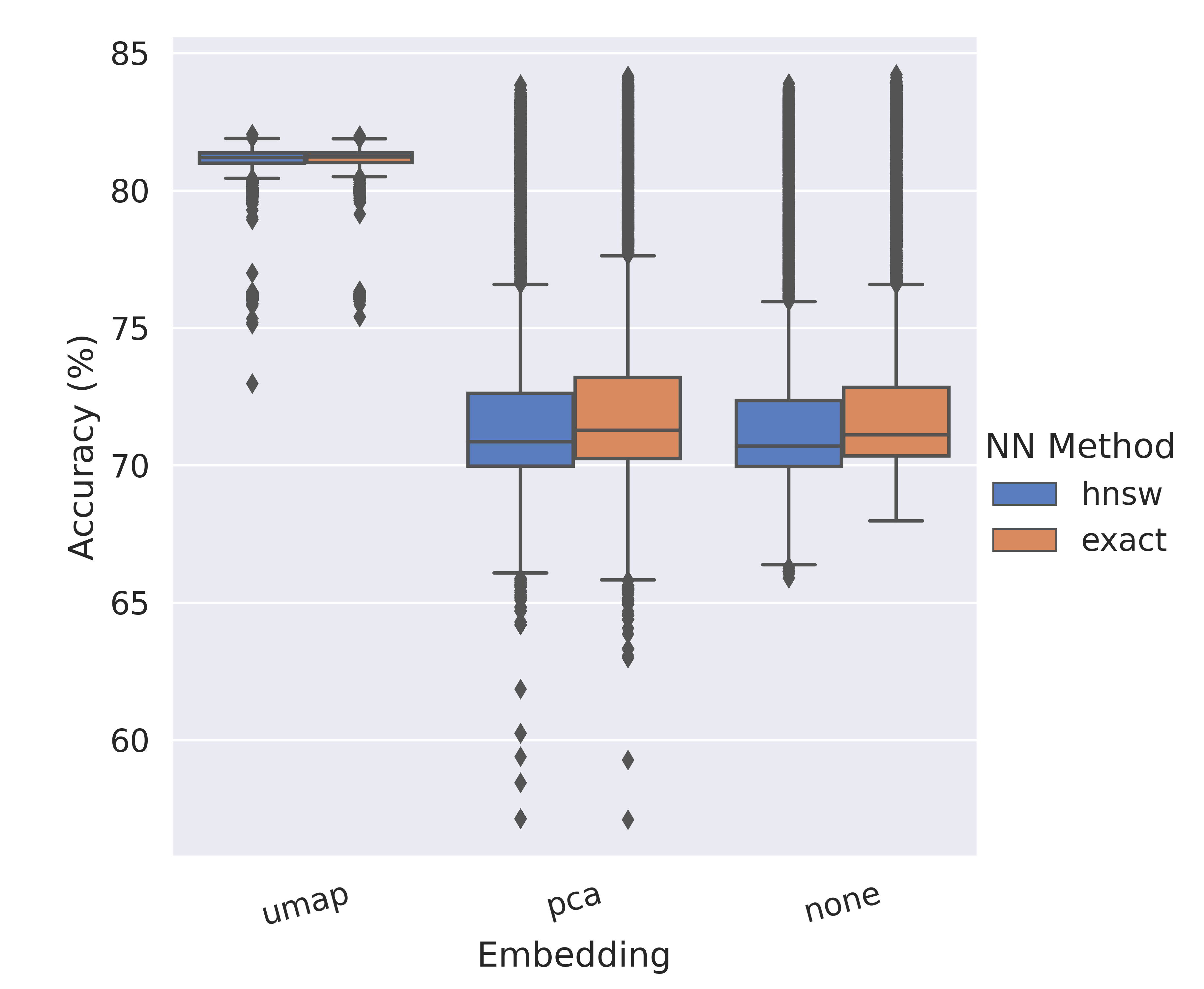}
    \caption{MuyGPs accuracies over embeddings for different nearest neighbors (NN) methods }
    \label{hnsw-vs-exact} 
\end{figure}

Our proposed machine learning model, MuyGPs, achieved competitive accuracy amongst state-of-the-art methods (refer to Table~\ref{tab:model_benchmark} for detailed comparisons). While the full Gaussian Process Classification (GPC) model demonstrates slightly superior accuracy, it is well-known for its computational complexity.
Figure~\ref{hnsw-vs-exact}  compares  the performance of MuyGPs using approximate nearest neighbor search method (hnsw) against the exact GPC on the ZTF dataset. While exact GPC achieves slightly higher accuracy, the gain is negligible compared to the significant difference in both train and test times. This makes MuyGPs a more practical choice for real-world applications involving large astronomical image datasets.

\begin{figure}[h!]
    \centering
    \includegraphics[scale=0.5]{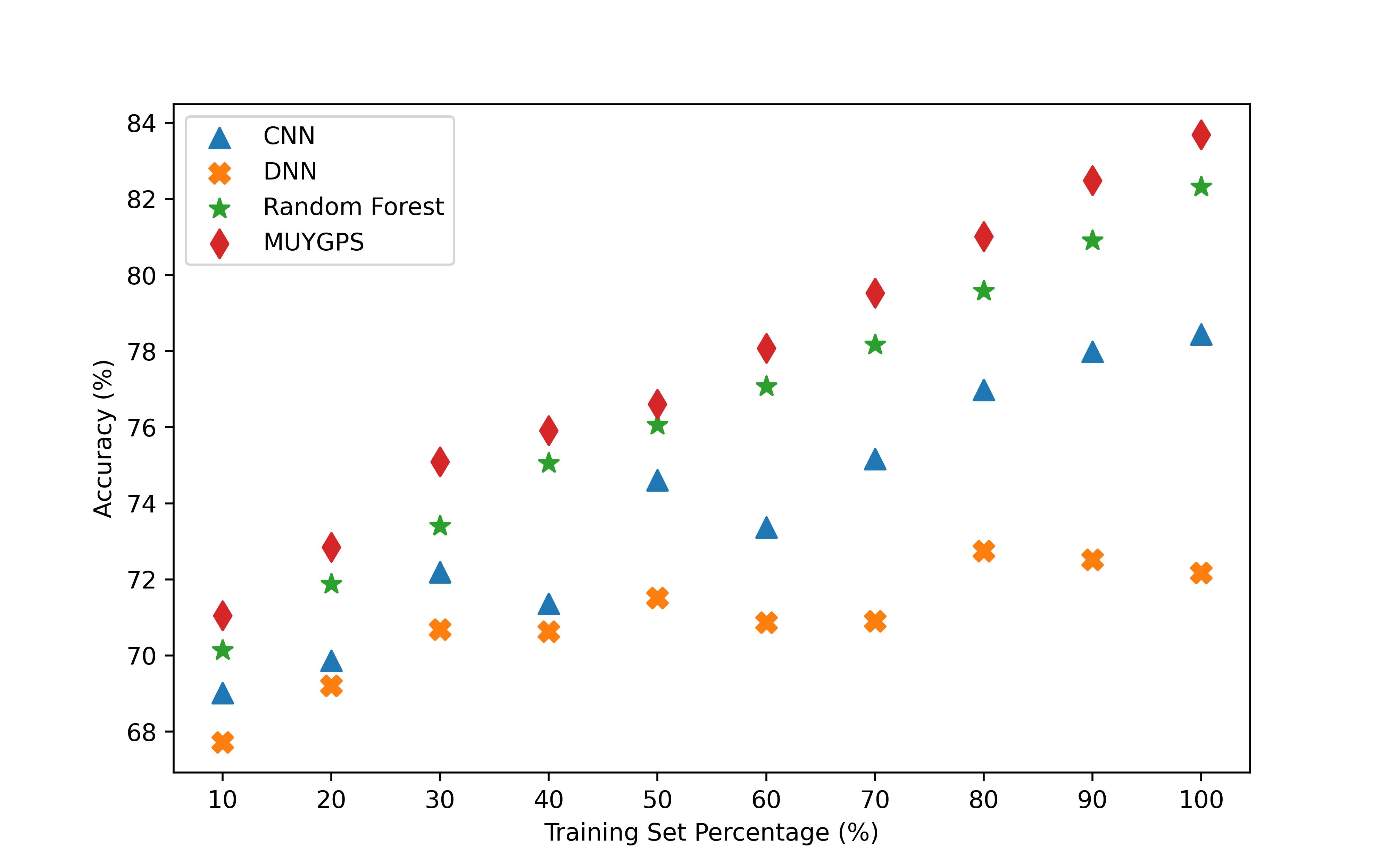}
    \caption{Accuracy for different amounts of training data.}
    \label{limited_data}
\end{figure}

\section{Discussion}
Our study underscores the importance of data normalization for optimal model performance. Figures~\ref{MuyGPs-top-umap-embedding} and Table~\ref{tab:model_benchmark} clearly demonstrate this effect. Even for visually similar images (Figure~\ref{images}), embedding techniques like supervised UMAP, when combined with normalization (e.g., local min-max), significantly improve class separation within the embedded space (Figure~\ref{raw-embedding}). This translates to better performance for all models, with MuyGPs consistently achieving superior accuracy (with or without normalization) compared to others (except exact GPC). However, the computational complexity of exact GPC makes it impractical for large-scale astronomical image classification tasks.
Our proposed MuyGPs model exhibits strong capability in distinguishing single stars from blended stars in low-resolution astronomical images. This is particularly evident in scenarios with limited data (Figure~\ref{limited_data}). This effectiveness might be partially attributed to MuyGPs' ability to handle noisy labels, a common characteristic of astronomical data, more effectively than other models.
By setting an uncertainty threshold, users can estimate expected accuracy and identify ambiguous data points requiring further human or specialized system intervention for annotation. This facilitates efficient analysis of large datasets by focusing resources on uncertain classifications.

This study successfully demonstrates the efficacy of the proposed automated pipeline, incorporating MuyGPs and data preprocessing techniques, in accurately classifying single and blended stars within low-resolution astronomical images. Furthermore, MuyGPs offers advantages in handling limited data and noisy labels, making it well-suited for real-world applications involving large astronomical image datasets.

\section{Conclusion}
We investigated a novel and cost-effective Gaussian Process  model (MuyGPs) to analyze star blended system in our universe in low-resolution astronomical images. MuyGPs model outperformed several popular image classification model such as CNN and random forests. The impact of data preprocessing and embedding are explored and discussed. Normalization techniques like local min-max scaling significantly improved the performance of all image classification models.
Embedding techniques, particularly supervised UMAP, in conjunction with normalization, further enhanced class separability and classification accuracy.
Our proposed MuyGPs model achieved superior accuracy compared to other models (except exact GPC) for this classification task. Additionally, MuyGPs demonstrated robustness in handling limited data and noisy labels, common characteristics of astronomical datasets.
The inherent uncertainty quantification capabilities of MuyGPs provides a valuable tool for identifying ambiguous data points requiring further investigation. This facilitates efficient and targeted analysis of large datasets, a valuable tool for real-world applications involving large astronomical surveys.

\section*{Supplementary Material}
Our codebase and the associated datasets used in this analysis are available in an open-source repository on GitHub: \url{https://github.com/cae0027/Stellar-Blends}.

\section*{Acknowledgement}
This work was performed under the auspices of the U.S. Department of Energy by Lawrence Livermore National Laboratory under Contract DE-AC52-07NA27344 with IM release number LLNL-JRNL-867196 and was supported by the LLNL-LDRD Program under Project No. 22-ERD-028.\\

\bibliographystyle{jds}
\bibliography{stellar-blends}

\begin{thebibliography}{25}
\providecommand{\natexlab}[1]{#1}

\bibitem[{Bosch et~al.(2018)Bosch, Armstrong, Bickerton, Furusawa, Ikeda, Koike et~al.}]{bosch2018hyper}
Bosch J, Armstrong R, Bickerton S, Furusawa H, Ikeda H, Koike M, et~al. (2018).
\newblock The hyper suprime-cam software pipeline.
\newblock \emph{Publications of the Astronomical Society of Japan}, 70(SP1): S5.

\bibitem[{Buchanan et~al.(2022)Buchanan, Schneider, Armstrong, Muyskens, Priest, and Dana}]{buchanan2022gaussian}
Buchanan JJ, Schneider MD, Armstrong RE, Muyskens AL, Priest BW, Dana RJ (2022).
\newblock Gaussian process classification for galaxy blend identification in lsst.
\newblock \emph{The Astrophysical Journal}, 924(2): 94.

\bibitem[{Challis et~al.(2015)Challis, Hurley, Serra, Bozzali, Oliver, and Cercignani}]{challis2015gaussian}
Challis E, Hurley P, Serra L, Bozzali M, Oliver S, Cercignani M (2015).
\newblock Gaussian process classification of alzheimer's disease and mild cognitive impairment from resting-state fmri.
\newblock \emph{NeuroImage}, 112: 232--243.

\bibitem[{Dey et~al.(2019)Dey, Schlegel, Lang, Blum, Burleigh, Fan et~al.}]{dey2019overview}
Dey A, Schlegel DJ, Lang D, Blum R, Burleigh K, Fan X, et~al. (2019).
\newblock Overview of the desi legacy imaging surveys.
\newblock \emph{The Astronomical Journal}, 157(5): 168.

\bibitem[{Gelfand et~al.(2010)Gelfand, Diggle, Guttorp, and Fuentes}]{gelfand2010handbook}
Gelfand AE, Diggle P, Guttorp P, Fuentes M (2010).
\newblock \emph{Handbook of spatial statistics}.
\newblock CRC press.

\bibitem[{Goumiri et~al.(2020)Goumiri, Muyskens, Schneider, Priest, and Armstrong}]{goumiri2020star}
Goumiri IR, Muyskens AL, Schneider MD, Priest BW, Armstrong RE (2020).
\newblock Star-galaxy separation via gaussian processes with model reduction.
\newblock \emph{arXiv preprint arXiv:2010.06094}.

\bibitem[{Hotelling(1933)}]{hotelling1933analysis}
Hotelling H (1933).
\newblock Analysis of a complex of statistical variables into principal components.
\newblock \emph{Journal of educational psychology}, 24(6): 417.

\bibitem[{Ivezi{\'c} et~al.(2019)Ivezi{\'c}, Kahn, Tyson, Abel, Acosta, Allsman et~al.}]{ivezic2019lsst}
Ivezi{\'c} {\v{Z}}, Kahn SM, Tyson JA, Abel B, Acosta E, Allsman R, et~al. (2019).
\newblock Lsst: from science drivers to reference design and anticipated data products.
\newblock \emph{The Astrophysical Journal}, 873(2): 111.

\bibitem[{Kim and Brunner(2016)}]{Kim_2016}
Kim EJ, Brunner RJ (2016).
\newblock Star{\textendash}galaxy classification using deep convolutional neural networks.
\newblock \emph{Monthly Notices of the Royal Astronomical Society}, 464(4): 4463--4475.

\bibitem[{Kim et~al.(2015)Kim, Brunner, and Carrasco~Kind}]{kim2015hybrid}
Kim EJ, Brunner RJ, Carrasco~Kind M (2015).
\newblock A hybrid ensemble learning approach to star--galaxy classification.
\newblock \emph{Monthly Notices of the Royal Astronomical Society}, 453(1): 507--521.

\bibitem[{Kim and Ghahramani(2006)}]{kim2006bayesian}
Kim HC, Ghahramani Z (2006).
\newblock Bayesian gaussian process classification with the em-ep algorithm.
\newblock \emph{IEEE Transactions on Pattern Analysis and Machine Intelligence}, 28(12): 1948--1959.

\bibitem[{McInnes et~al.(2018)McInnes, Healy, and Melville}]{McInnes_2018}
McInnes L, Healy J, Melville J (2018).
\newblock Umap: Uniform manifold approximation and projection for dimension reduction.

\bibitem[{{Monthly Updates }(2024)}]{rubin_2024}
{Monthly Updates } (2024).
\newblock {Large Synoptic Survey}.
\newblock \url{https://www.lsst.org/about/project-status}.
\newblock [Online; accessed 16-March-2024].

\bibitem[{Muyskens et~al.(2021{\natexlab{a}})Muyskens, Priest, Goumiri, and Schneider}]{muyskens2021MuyGPs}
Muyskens A, Priest B, Goumiri I, Schneider M (2021{\natexlab{a}}).
\newblock Muygps: Scalable gaussian process hyperparameter estimation using local cross-validation.
\newblock \emph{arXiv preprint arXiv:2104.14581}.

\bibitem[{Muyskens et~al.(2021{\natexlab{b}})Muyskens, Priest, Goumiri, and Schneider}]{MuyGPs2021git}
Muyskens A, Priest BW, Goumiri I, Schneider M (2021{\natexlab{b}}).
\newblock Fast implementation of the muygps scalable gaussian process algorithm.
\newblock https://github.com/LLNL/MuyGPyS.

\bibitem[{Muyskens et~al.(2022)Muyskens, Goumiri, Priest, Schneider, Armstrong, Bernstein et~al.}]{muyskens2022star}
Muyskens AL, Goumiri IR, Priest BW, Schneider MD, Armstrong RE, Bernstein J, et~al. (2022).
\newblock Star--galaxy image separation with computationally efficient gaussian process classification.
\newblock \emph{The Astronomical Journal}, 163(4): 148.

\bibitem[{Nickisch and Rasmussen(2008)}]{nickisch2008approximations}
Nickisch H, Rasmussen CE (2008).
\newblock Approximations for binary gaussian process classification.
\newblock \emph{Journal of Machine Learning Research}, 9(Oct): 2035--2078.

\bibitem[{Odewahn et~al.(1992)Odewahn, Stockwell, Pennington, Humphreys, and Zumach}]{odewahn1992automated}
Odewahn S, Stockwell E, Pennington R, Humphreys RM, Zumach W (1992).
\newblock Automated star/galaxy discrimination with neural networks.
\newblock In: \emph{Digitised Optical Sky Surveys: Proceedings of the Conference on ‘Digitised Optical Sky Surveys’, Held in Edinburgh, Scotland, 18--21 June 1991}, 215--224. Springer.

\bibitem[{Rasmussen et~al.(2006)Rasmussen, Williams et~al.}]{rasmussen2006gaussian}
Rasmussen CE, Williams CK, et~al. (2006).
\newblock \emph{Gaussian processes for machine learning}, volume~1.
\newblock Springer.

\bibitem[{Sanchez et~al.(2021)Sanchez, Mendoza, Kirkby, Burchat et~al.}]{sanchez2021effects}
Sanchez J, Mendoza I, Kirkby DP, Burchat PR, et~al. (2021).
\newblock Effects of overlapping sources on cosmic shear estimation: Statistical sensitivity and pixel-noise bias.
\newblock \emph{Journal of Cosmology and Astroparticle Physics}, 2021(07): 043.

\bibitem[{Scheirer et~al.(2012)Scheirer, de~Rezende~Rocha, Sapkota, and Boult}]{scheirer2012toward}
Scheirer WJ, de~Rezende~Rocha A, Sapkota A, Boult TE (2012).
\newblock Toward open set recognition.
\newblock \emph{IEEE transactions on pattern analysis and machine intelligence}, 35(7): 1757--1772.

\bibitem[{Sevilla-Noarbe and Etayo-SotosI(2015)}]{Sevilla2015}
Sevilla-Noarbe I, Etayo-SotosI P (2015).
\newblock Effect of training characteristics on object classification: An application using boosted decision trees.
\newblock \emph{Astronomy and Computing}, 11: 64--72.

\bibitem[{Stein(1999)}]{stein1999interpolation}
Stein ML (1999).
\newblock \emph{Interpolation of spatial data: some theory for kriging}.
\newblock Springer Science \& Business Media.

\bibitem[{Van~der Maaten and Hinton(2008)}]{van2008visualizing}
Van~der Maaten L, Hinton G (2008).
\newblock Visualizing data using t-sne.
\newblock \emph{Journal of machine learning research}, 9(11).

\bibitem[{Yang et~al.(2021)Yang, Zhou, Li, and Liu}]{yang2021generalized}
Yang J, Zhou K, Li Y, Liu Z (2021).
\newblock Generalized out-of-distribution detection: A survey.
\newblock \emph{arXiv preprint arXiv:2110.11334}.

\end{thebibliography}

\begin{appendix}
\section{Appendix section}

\begin{table}[h!]
  \centering
  \begin{tabular}{|l|l|}
    \hline
    \textbf{Aspect}                & \textbf{Description}                                \\
    \hline
    \textbf{Input Layer}           & Input Shape: (100 dimensions)                      \\
                                   & Preprocessing: Data Normalizations in sec \ref{data-norm}                      \\
    \textbf{Hidden Layers}         & Layer 1: Linear, Neurons: 70, Activation: ReLU     \\
                                   & Layer 2: Linear, Neurons: 64, Activation: ReLU     \\
                                   & Layer 3: Linear, Neurons: 32, Activation: ReLU     \\
    \textbf{Output Layer}          & Linear, Neurons: 1, Activation: Sigmoid            \\
    \textbf{Training Details}      & Loss Function: BCELoss                            \\
                                   & Optimizer: Adam                                   \\
                                   & Learning Rate: 0.0001                             \\
                                   & Batch Size: 128                                   \\
                                   & Number of Epochs: 1000 \\
    \textbf{Model Size}            & Total Parameters: $13, 727$          \\
    \textbf{Bias Setting}          & Bias: True in all layers                          \\
    \textbf{Software} & Framework:  PyTorch                        \\

    \hline
  \end{tabular}
  \caption{Deep neural network model architecture and training details}
  \label{dnn-model}
\end{table}

\begin{table}[h!]
  \centering
  \begin{tabular}{|l|l|}
    \hline
    \textbf{Aspect}                & \textbf{Description}                                \\
    \hline
    \textbf{Input Layer}           & Input Dim: $10 \times 10$                             \\
    \textbf{Convolutional Layers}  & Layer 1: Conv2D(1,32); Layer2: Conv2D(32, 64)\\
                                    & kernel: $3\times 3$, stride: (1,1), padding:(1,1)\\        
                                    & Activation: ReLU \\
    \textbf{Fully Connected Layers}& fc1:Linear(6400, 64), ReLU               \\
                                    & fc2: Linear(64, 2)\\
    \textbf{Training Details}      & Loss Function: CrossEntropyLoss                             \\
                                   & Optimizer: Adam                                   \\
                                   & Learning Rate: 0.001                             \\
                                   & Batch Size: 64                                   \\
                                   & Number of Epochs: 10                            \\
    \textbf{Model Size}            & Total Parameters: $428,610$                        \\
    \textbf{Bias Setting}          & Bias: True in fc1, fc2 layers                    \\
    \textbf{Software}              & Framework: PyTorch                                 \\
    \hline
  \end{tabular}
  \caption{Convolutional neural network model architecture and training details}
  \label{cnn-model}
\end{table}

\begin{figure}[h!]
    \centering
    \includegraphics[scale=0.39]{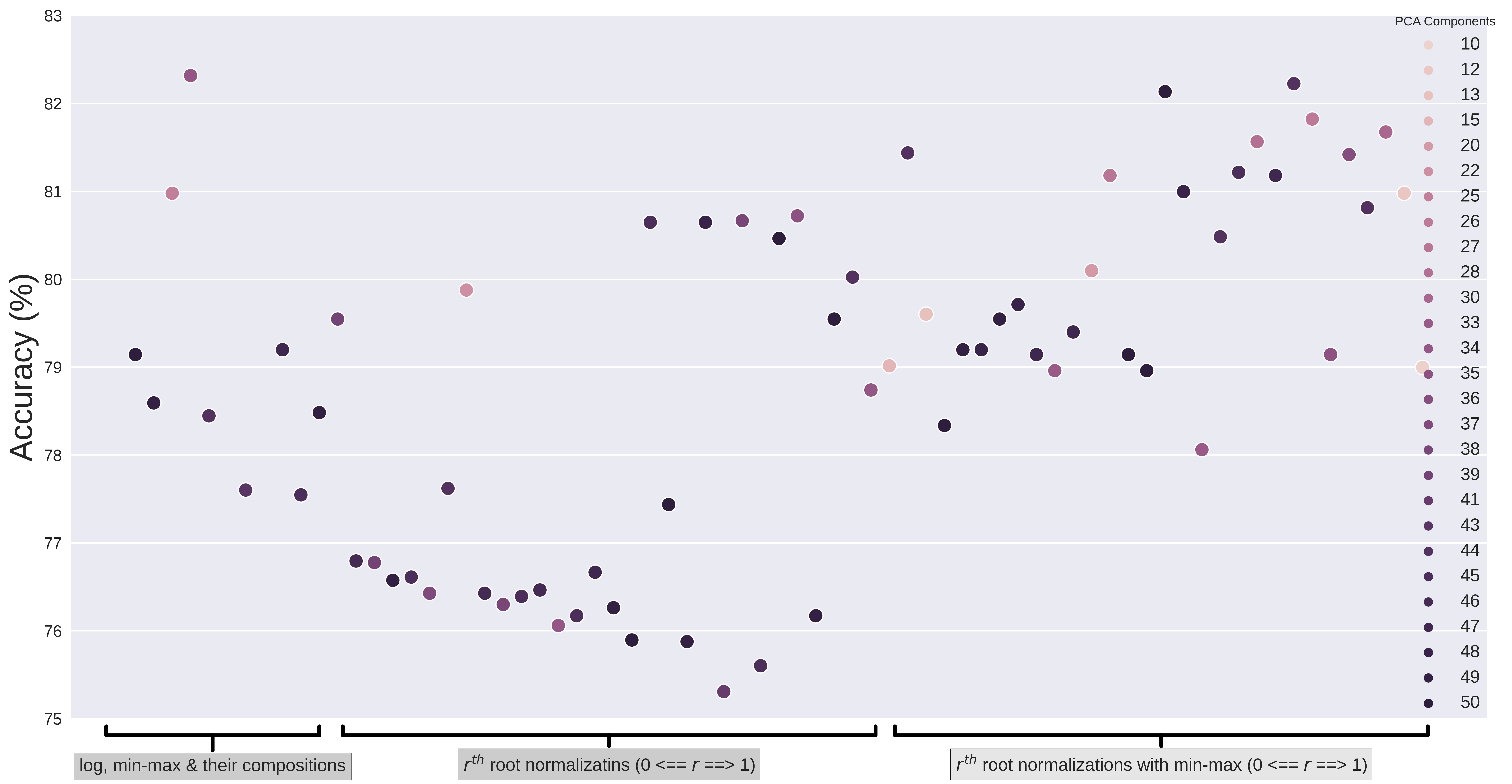}
    \caption{MuyGPs performance on outputs from PCA embedding for different normalziations. 
    }
    \label{MuyGPs-pca-embedding} 
\end{figure}

\begin{figure}[h!]
    \centering
    \includegraphics[scale=0.39]{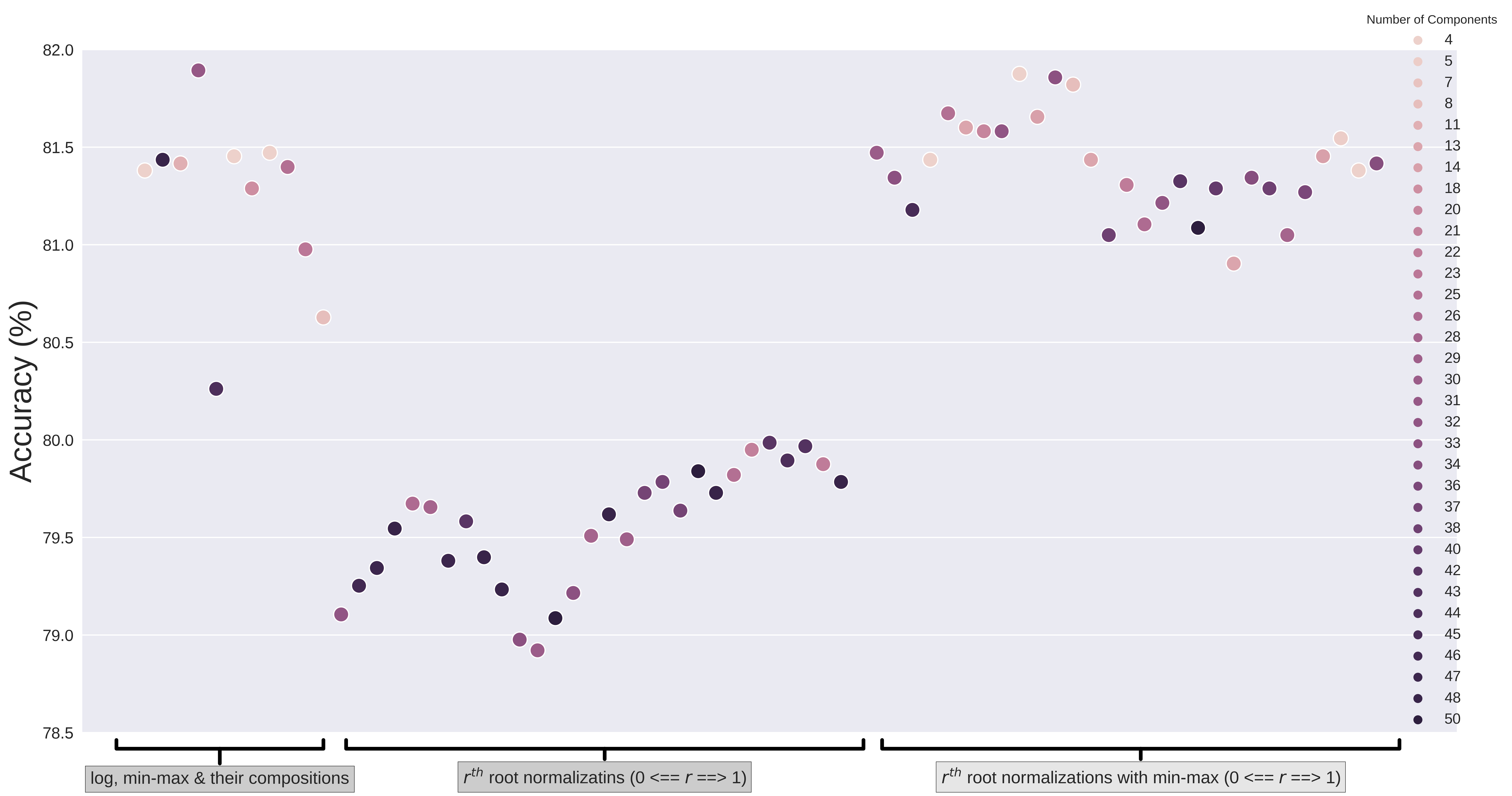}
    \caption{MuyGPs performance on outputs from supervised UMAP embedding for different normalizations. 
    }
    \label{MuyGPs-umap-embedding} 
\end{figure}

\end{appendix}
\end{document}